\documentclass[aps,twocolumn,nofootinbib]{revtex4}

\usepackage{graphicx}
\usepackage{amssymb, amsmath,mathtools}
\usepackage{bbm}
\usepackage{enumerate}

\graphicspath{{./}{./plots/}}

\usepackage{xcolor}

\usepackage{bm}   %bold math

\usepackage{hyperref} % add hypertext capabilities
\hypersetup{
colorlinks = true,   %Colours links instead of ugly boxes
urlcolor = blue,    %Colour for external hyperlinks
linkcolor = magenta,   %Colour of internal links
citecolor = blue  %Colour of citations
}
\usepackage{float}
\usepackage{color}
\begin{document}

\title{Superconductivity in Luttinger semimetals near the SU(4) limit}

\author{Majid Kheirkhah and Igor F. Herbut}

\affiliation{ Department of Physics, Simon Fraser University, Burnaby, British Columbia V5A 1S6, Canada}

\begin{abstract}
We consider the spin-3/2 Luttinger fermions with contact attraction near the SU(4)-symmetric limit of vanishing Luttinger spin-orbit-coupling parameter responsible for band inversion, and at finite chemical potential. In the case of exact SU(4) symmetry the previously considered $s$-wave and five $d$-wave superconducting order parameters together form a six-dimensional irreducible representation which transforms as an antisymmetric tensor under SU(4). In this limit, we find the SU(4) [$\simeq$ SO(6)] symmetry to be spontaneously broken down to the SO(5) at the superconducting transition. When the SU(4) is reduced to the SO(3) by the weak band-inverting kinetic energy term, we show that at low temperatures the superconducting state is $is+d$, with a dominant $s$ and a small $d$ component, and spontaneously broken time-reversal symmetry. Relevance to superconductivity in doped semiconductors with diamond structure is discussed.
\end{abstract}

\maketitle

\section{Introduction}

Superconductivity in spin-orbit coupled materials such as three-dimensional Luttinger semimetals \cite{luttinger} of electrons with effective spin-3/2 has received plenty of attention lately \cite{butch, bay, kim}. When the spin-orbit coupling is strong so that the system exhibits band inversion, an attractive interaction can lead to various superconducting states  with Cooper pairs with total angular momentum $j=0,1,2,3$
\cite{boettcher1, meinert, brydon, savary, boettcher2, venderbos, roy, mandal, link, Ipsita}. The opposite limit of weak spin-orbit coupling without the band inversion, relevant to semiconductors with diamond and zinc blende structure, for example \cite{ghorashi, cardona}, has in contrast not been discussed as much, in spite of reports of superconductivity at low temperatures \cite{ekimov}. In this limit the usual SO(3) rotational symmetry of the Luttinger Hamiltonian is close to being enlarged to the maximal SU(4) symmetry that four-component fermions may have, and some of the previously studied superconducting states may belong to the same irreducible representation of the larger symmetry group \cite{herbutfisherbook, herbutmandal}. In this paper, we consider the simplest such situation, when an attractive density-density interaction, being itself SU(4) symmetric, does not discriminate between the (single, $j=0$) $s$-wave and the (five, $j=2$) $d$-wave pairings of Luttinger fermions. The six complex order parameters are shown to form the irreducible representation (irrep) which transforms as the antisymmetric tensor under SU(4). Since SU(4) = Spin(6), i.e., SU(4) is the spin version of SO(6) \cite{zee}, the six order parameters, appropriately defined, may equivalently be considered to form the vector representation of SO(6).

We first derive the Ginzburg-Landau free energy for the antisymmetric tensor order parameter in the SU(4)-symmetric limit of vanishing spin-orbit coupling, and find that the quartic term(s) dictate that the symmetry at the superconducting transition is spontaneously broken to the SO(5). Interestingly, at the same time the breaking of the time-reversal symmetry remains undecided. When we allow a weak band-inverting term in the Luttinger Hamiltonian that reduces the SU(4) to the usual rotational SO(3), however, we find that the $s$-wave component generically ends up with a higher critical temperature and therefore inevitably condenses first, with the $d$-wave components to follow at a lower temperature. Interestingly, the ratio of the subdominant $d$-wave and the dominant $s$-wave transition temperatures is found to be given by a number which is universal to the leading order in weak band-inverting Luttinger parameter. Most importantly, the $d$-wave components have their common overall phase differ from the $s$-wave component by $\pi/2$, and the time reversal is consequently broken at low temperatures.

The paper is organized as follows. In sec.~II, we introduce the noninteracting Luttinger fermions and establish notation. In sec.~III, we add attractive interaction between Luttinger fermions and define the six relevant superconducting order parameters. In sec.~IV A we first derive the Ginzburg-Landau free energy in the strict SU(4) limit. Sec.~IV B discusses how the time-reversal symmetry acts on the order parameters. In sec.~IV C the Ginzburg-Landau free energy is derived at weak spin-order-coupling Luttinger parameter, and the additional superconducting transition within the $s$-wave phase is discussed. Finally, we summarize our results and discuss relevance to doped semiconductors in the last section. Calculational details are relegated to five Appendixes.

%==================
\section{Luttinger Hamiltonian}

The low-energy action for interacting Luttinger fermions can be written as $S = S_0 + S_{\rm int}$. The noninteracting action $S_0$ is
\begin{align}
S_0 =  \int \frac{d^3 \bm{p}}{(2\pi)^3}
\int_0 ^{\beta}
\hspace{-1mm} d\tau ~
\psi^{\dagger}_{\bm{p}} (\tau)
\Big[ \partial_{\tau} + \mathcal{H}_0(\bm{p}) \Big]
\psi_{\bm{p}} (\tau),
\label{Lag_0}
\end{align}
and $\psi_{\bm{p}} (\tau) = (c_{\bm{p}, \frac{3}{2}}, ~c_{\bm{p}, \frac{1}{2}},~c_{\bm{p}, -\frac{1}{2}},~c_{\bm{p}, -\frac{3}{2}})^{\rm T}$ is the four-component Grassmann field, $\beta = 1/k_B T$ is the inverse of temperature $T$, $k_B$ is the Boltzmann constant, and $\tau$ represents imaginary time. The single-particle normal state Luttinger Hamiltonian reads
\begin{equation}
\mathcal{H}_0(\bm{p}) = (p^2 - \mu) 1_{4\times4} + \lambda \sum^5_{a = 1} d_a(\bm{p}) \gamma_a,
\label{Hami_normal}
\end{equation}
where $\bm{p} = (p_1,p_2,p_3)$ is the momentum, $\lambda$ measures the strength of spin-orbit coupling and is a real ``band-inversion parameter", and $\mu$ is the chemical potential. The five Hermitian matrices $\gamma_a$ obey the Clifford algebra $\{\gamma_a, \gamma_b \} = 2\delta_{ab}$ and will here be chosen to be $\gamma_1 = \sigma_1 \otimes  1_{2 \times 2}$, $\gamma_2 = \sigma_3 \otimes \sigma_3$,
$\gamma_3 = \sigma_3 \otimes \sigma_1$, $\gamma_4 = \sigma_3 \otimes \sigma_2$,
$\gamma_5 = \sigma_2 \otimes 1_{2\times2} $, where $\sigma_i$ $i=1,2,3$ are the usual Pauli matrices. The five real ($l = 2$) spherical harmonics $d_i(\bm{p})$ are defined as
\begin{align}
d_1(\bm{p}) &= \frac{\sqrt{3}}{2}(p^2_x - p^2_y),
\hspace{5mm}
d_2(\bm{p}) = \frac{1}{2}(3p^2_z - p^2),
\nonumber
\\
d_3(\bm{p}) &= \sqrt{3} p_x p_z,
\hspace{1.5mm}
d_4(\bm{p}) = \sqrt{3} p_y p_z,
\hspace{1.5mm}
d_5(\bm{p}) = \sqrt{3} p_x p_y.
\nonumber
\end{align}
We assumed full rotational symmetry, since often the additional terms that reduce the rotational symmetry to cubic symmetry are weak \cite{cardona}, as well as  for reasons of simplicity.  In addition, it is known that one effect of long-range Coulomb interaction is to make the single-particle dispersion progressively more isotropic with lowering of the energy \cite{abrikosov, moon, herbut1, janssen1, boettcher3}.
\begin{figure}
\centering
\includegraphics[scale=0.57]{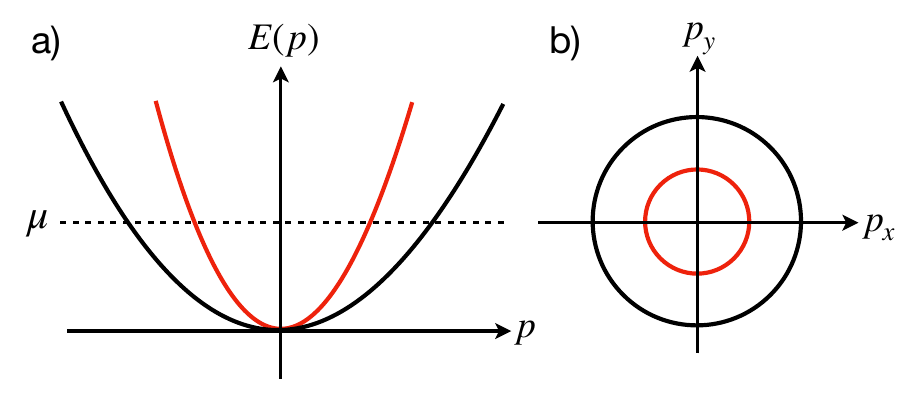}
\caption{(Color online) a) Schematic band structure of the Luttinger Hamiltonian for $0<\lambda<1$ where the two upward doubly degenerate bands cross at the $\bm{p}=0$ point. b) The two-dimensional projection of the two three-dimensional Fermi surfaces, with the quasiparticles with the third component of the spin either $\pm 1/2$ (black line) or $\pm 3/2$ (red line).}
\label{fig_a}
\end{figure}

The eigenvalues of the Luttinger  Hamiltonian read as
\begin{equation}
E_{\pm}(p) = p^2(1\pm \lambda) - \mu,
\end{equation}
and exhibit a quadratic touching point of all four dispersion bands at $\bm{p}=0$, that is the zone center. Away from the zone center the fourfold degeneracy is reduced to twofold when $\lambda\neq 0 $. The remaining degeneracy is  due to the presence of both inversion and time reversal. The time-reversal operator in our representation is given by  $\mathcal{T}= \gamma_{45} \mathcal{K}$, where $\gamma_{45} = i \gamma_4 \gamma_5$, and $\mathcal{K}$ denotes the complex conjugation \cite{boettcher1}. When the Luttinger parameter $|\lambda|>1$, two of the bands are dispersing upwards and the other two downwards, and the bands become ``inverted". Here we will be interested in the opposite limit of $|\lambda| <1$ when both pairs of bands are dispersing the same way [for simplicity chosen to be upward in Fig.~\ref{fig_a}(a)], and all four bands intersect the assumed finite chemical potential [Fig.~\ref{fig_a}(b)]. When $\lambda=0$, the degeneracy is fourfold at all momenta and the Hamiltonian becomes fully SU(4) symmetric.

%==============
\section{Pairing interaction}

We assume next the interacting part of the action $S_{\rm int}$ to be given by a simple model
\begin{align}
S _{\rm int} = -g
\int  d^3 \bm{x}
\int_0^{\beta} \hspace{-1mm} d\tau
 \big[ \psi^{\dagger}_{\bm x} (\tau)  \psi_{\bm x} (\tau)
 \big]^2,
\label{Lag_int}
\end{align}
with $g>0$ and $\bm{x}=(x_1, x_2, x_3)$ as a coordinate, which corresponds to the density-density attractive contact interaction. The interaction term is clearly invariant under $\psi \to U \psi$, with $U \in$ SU(4). We omitted the second independent contact  term that would reduce the symmetry of $S_{\rm int}$ to SO(3) \cite{herbut1}. This way the only reduction of SU(4) in the action comes from the kinetic energy term when the spin-orbit parameter $\lambda$ is finite.

One may use the Fierz identity \cite{boettcher1, herbut2} and decompose the above interaction into $s$- and $d$-wave pairing channels:  $ 4 [\psi^{\dagger}_{\bm x} (\tau)  \psi_{\bm x} (\tau)  ]^2 = \mathcal{L}_s + \mathcal{L}_d $,  where
\begin{align}
\mathcal{L}_s &= \big[\psi^{\dagger}_{\bm x} (\tau) \gamma_{45} \psi^{\ast} _{\bm x} (\tau) \big]
\big[ \psi^{\rm T}_{\bm x} (\tau) \gamma_{45} \psi_{\bm x} (\tau) \big],
\\
\mathcal{L}_d &= \sum_{a = 1}^5
\big[ \psi^{\dagger}_{\bm x} (\tau) \gamma_a \gamma_{45} \psi^{\ast}_{\bm x} (\tau) \big]
\big[ \psi^{\rm T}_{\bm x} (\tau) \gamma_{45} \gamma_a \psi_{\bm x} (\tau) \big].
\end{align}
$\mathcal{L}_s$ is the $s$-wave pairing term, with the $s$-wave as a singlet under the rotational SO(3) symmetry. $\mathcal{L}_d$ is the $d$-wave pairing term \cite{boettcher1} with five $d$-wave components which under rotational SO(3) transform as a symmetric irreducible tensor \cite{boettcher2}. In our simple model both terms evidently come with the same pairing interaction $-g/4$.

One may define all six complex order parameters together as
\begin{equation}
\Delta_a = \frac{g}{4} \big \langle \psi_{\bm p}^{\rm T}(\omega)
\gamma_{45} \mathcal{A}_a
 \psi_{-\bm p}(-\omega)  \big\rangle,
\end{equation}
with the matrices $\mathcal{A}_a = \{i1_{4 \times 4},\gamma_1,\gamma_2,\dots,\gamma_5\}$ for $a = 0,1,\dots,5$. Note the imaginary unit included in the definition of the first ($s$-wave) component. The six components may be understood as specifying the antisymmetric four-dimensional matrix order parameter, which in our notation can be written as
\begin{align}
\phi &= \sum_{a=0}^5 \Delta_a \mathcal{A}_a \gamma_{45}
\nonumber
\\
&=
\begin{pmatrix}
 0 & -\Delta_5-i \Delta_1 & \Delta_4+i \Delta_3 & \Delta_0-i \Delta_2 \\
 \Delta_5+i \Delta_1 & 0 & -\Delta_0-i \Delta_2 & \Delta_4-i \Delta_3 \\
 -\Delta_4-i \Delta_3 & \Delta_0+i \Delta_2 & 0 & \Delta_5-i \Delta_1 \\
 -\Delta_0+i \Delta_2 & -\Delta_4+i \Delta_3 & -\Delta_5+i \Delta_1 & 0
\end{pmatrix}.
\label{main_phi}
\end{align}
Under the SU(4) transformation $\psi \to U \psi$ the matrix $\phi$ transforms as $\phi \to U^{\rm T} \phi U $, and therefore remains antisymmetric. Six complex components, the $s$-wave and the five $d$-waves together, transform therefore as the irrep ``6" of the SU(4), i.e., as the antisymmetric tensor.

The Lie group SU(4) represents, on the other hand, the spin version of the SO(6) \cite{zee}. Indeed, one can think of the Lie algebra of SO(6) as being closed by the five matrices $\gamma_1, \gamma_2,\dots,\gamma_5$, and the remaining ten $\gamma_{ab}=i\gamma_a \gamma_b $, $a<b$, with the former set being a vector under the latter set, which alone closes the Lie algebra SO(5).  The antisymmetric tensor irrep must therefore correspond to some irrep of SO(6). It is straightforward to check that with the $s$-wave component defined as above, that is with the imaginary unit included, the SU(4) transformations generated by $\gamma_a$ for $a=1,2,..., 5$ rotate in the ``0a" plane, whereas those generated by $\gamma_{ab}$ rotate in the ``ab" plane (see Appendix \ref{apend_rotation}). In other words, the order parameter as defined above is a simple vector under SO(6).

%==================
\section{Ginzburg-Landau theory}

\subsection{SU(4) limit when $\lambda = 0$}

We now proceed to integrate out the Luttinger fermions and derive the Ginzburg-Landau free energy for the superconducting order parameters. We do this first for $\lambda=0$, up to quartic terms, and assuming the order parameter to be uniform. The SU(4) and the standard particle-number U(1) symmetries dictate the form of the Ginzburg-Landau free energy in general to be
\begin{align}
\hspace{-2mm}
F( \phi ) &= a~ {\rm tr}(\phi \phi^{\dagger}) +
b_1 \big[ {\rm tr}(\phi \phi^{\dagger}) \big]^2
+ b_2~ {\rm tr}(\phi \phi^{\dagger} \phi \phi^{\dagger})
\nonumber
\\&= 4a \Delta ^\dagger \Delta + (16b_1 + 8b_2) (\Delta ^\dagger \Delta )^2
-4 b_2 |\Delta ^{\rm T} \Delta|^2,
\label{GL_su4}
\end{align}
where $\Delta= (\Delta_0, \Delta_1, ..., \Delta_5)$. The coefficients $a$, $b_1$, and $b_2$ to the leading order at low temperatures are found to be
\begin{align}
a &=
\frac{1}{g}
-\frac{\mathcal{N}(\mu,0)}{2}
(\log \frac{\Omega}{T} + C),
\label{eq_log}
\\
b_1 &= 0,
\\
b_2 &=
\frac{\zeta\mathcal{N}(\mu,0)}{2T ^2},
\label{eq_b2_main}
\end{align}
where $\mathcal{N}(\mu,0) = \sqrt{\mu} / \pi^2 $ is the density of states at the Fermi level for $\lambda =0$, $\zeta = 0.2131$ is a constant, $\Omega$ is the usual ultraviolet cutoff, and the constant $C=\log (2e^{\gamma}/\pi) $ (see Appendixes \ref{append_loop} and \ref{append_lambda_zero}). In the weak coupling limit $g \ll 1$, the coefficient $a$ changes sign at the critical temperature
\begin{align}
T_c = \Omega e^{-\dfrac{2}{g \mathcal{N}(\mu,0)}}.
\label{tc1}
\end{align}

Since the coefficient $b_2$ is positive the resulting superconducting order maximizes the last term in Eq.~(\ref{GL_su4}), which means that, modulo an overall phase,  $\Delta = |\Delta| \hat{n}$, with $\hat{n}$ as a real six-component unit vector, $\hat{n}^{\rm T} \hat{n} =1$, and $|\Delta| = \sqrt{ -a/(2b_2)} $ for $a<0$. The ground state is invariant under SO(5) transformations that leave the  unit vector $\hat{n}$ invariant. It is also easy to check that the quasiparticle spectrum has a full isotropic gap $|\Delta|$.
%=================
\subsection{Time-reversal symmetry}

Under time reversal the fermion field transform as $\psi \to \gamma_{45} \psi^* $, and $ \psi^{\rm T} \to \psi^{\dagger} \gamma_{45}^{\rm T}$. Since $\gamma_{45}^{\rm T} = -\gamma_{45}$, one readily finds that under time reversal
\begin{align}
\Delta_a \to s \Delta_a ^* ,
\end{align}
with the sign $s= +$ for $a=0$, and $s=-$ for $a=1,2,..., 5$. The difference in the transformation property between the $s$- and $d$-wave components stems from the imaginary unit that was included in the definition of $\Delta_0$, and which was necessary for the six-component $\Delta$ to be a vector under SO(6). This means that if the six-component superconducting order parameter has both the $s$-wave and some of the $d$-wave components finite the time-reversal symmetry is broken; otherwise it is not. The free energy of the superconducting configuration, however, in either case is the same.
%=================
\subsection{ $\lambda\neq 0$ }

Let us now allow a finite but small spin-orbit parameter $\lambda$ and rederive the Ginzburg-Landau free energy by integrating out the fermions. The symmetry is now only the standard rotational SO(3), so the six-dimensional irrep of SU(4) is broken into $s$-wave singlet $\Delta_s= \Delta_0$ and $d$-wave quintuplet $\Delta_d= (\Delta_1, ...,  \Delta_5)$, which are two distinct irreps of SO(3). The Ginzburg-Landau free energy up to the quartic terms now reads
\begin{align}
\hspace{-1mm}
F(\Delta_s,\Delta_d) &= a_1 |\Delta_s|^2 + a_2 \Delta_d ^\dagger \Delta_d
+ q_1 |\Delta_s |^4
+ q_2 (\Delta_d ^\dagger \Delta_d) ^2
\nonumber
\\&\quad
+ q_3 |\Delta_d ^{\rm T} \Delta_d|^2
+ q_4 |\Delta_s|^2  \Delta_d ^\dagger \Delta_d
\nonumber
\\&\quad
+ q_5 (\Delta^{\ast 2}_s \Delta_d ^{\rm T} \Delta_d  +  \Delta^{2}_s \Delta_d ^\dagger \Delta_d ^* )
\nonumber
\\&\quad
+ q_6 \Delta^{\ast}_s {\rm tr} (M^2 M^{\ast} ) +  q^{\ast}_6 \Delta _s {\rm tr} (M^{\ast 2} M )
\nonumber
\\&\quad
+ q_7 {\rm tr}(M M^{\ast} M M^{\ast}),
\label{eq_quartic1}
\end{align}
where we found it convenient to group the five-component $\Delta_d$ into a three-dimensional symmetric traceless matrix $M$, as
\begin{equation}
M = \sum_{i =1}^5  \Delta_i M_i,
\end{equation}
and $M_i$ ($i=1,2,\dots,5$) are five real symmetric traceless $3 \times 3$ Gell-Mann matrices that provide a basis (see Appendix \ref{append_lambda_finite}).  The matrix $M$ transforms then as a symmetric irreducible tensor under SO(3), $M\to O^{\rm T} M O$, with $O$ in the fundamental representation. Equation~(\ref{eq_quartic1}) represents the most general fourth-order expression that is invariant under SO(3) and common global gauge transformation of the $s$- and $d$-wave order parameters.

The quadratic coefficient for the $s$-wave remains essentially unchanged (see Appendix \ref{append_lambda_finite}):
\begin{align}
a_1 &=  \frac{4}{g}
- 2 \mathcal{N}(\mu,\lambda) ( \log \frac{\Omega}{T} + C ),
\label{eq_a11}
\end{align}
where only the density of states at the Fermi level is modified into
\begin{equation}
\mathcal{N} (\mu,\lambda) =
\sum_{\eta = \pm}  \mathcal{N}_{\eta} (\mu,\lambda)
=
\frac{\sqrt{\mu}}{2 \pi^2}
\sum_{\eta = \pm} \frac{1}{ (1+\eta \lambda)^{3/2}}.
\label{DOS}
\end{equation}
For the $d$-wave components, on the other hand, we find the quadratic coefficient to be
\begin{align}
a_2 = a_1 + \frac{8 \zeta}{5} \mathcal{N}(\mu,0) (\frac{\lambda \mu}{T})^2
+ \mathcal{O}((\lambda \mu)^4),
\label{maintext_diff_a_2_1}
\end{align}
to the leading order in small parameter $\lambda$, the relevant criterion being defined as $\lambda \mu/T  \ll 1$ (Appendix \ref{append_lambda_finite}). When $\lambda=0$, we thus recover $a_2 = a_1$. Remarkably, at finite $\lambda$, one finds $a_2 > a_1$ for all temperatures, and therefore at the critical temperature
\begin{align}
T_c^s = \Omega e^{-\dfrac{2}{g \mathcal{N}(\mu,\lambda)}},
\label{tcs}
\end{align}
only the $s$-wave component condenses.

Furthermore, the Ginzburg-Landau coefficient $q_7 =0$ exactly at all values of $\lambda$. To the zeroth order in $\lambda$ (i.e., when $\lambda=0$) by matching Eqs.~(\ref{GL_su4}) and (\ref{eq_quartic1}) we also find that $q_6=0$, and that (Appendix \ref{append_lambda_finite})
\begin{align}
4b_2 = q_1 = \frac{q_2}{2} = -q_3 = \frac{q_4}{4} = -q_5.
\end{align}
Right below $T_c^s$, $\Delta_s  = \sqrt{-a_1/(2 q_1)}$, and may be chosen real, and $\Delta_d =0$. We find that the cubic coefficient $q_6 \sim \lambda^3 $ and thus negligible \cite{sim}. Since to the leading order in small $\lambda$ the coefficient $q_3 $ is still negative, the resulting $d$-wave order parameter $\Delta_d$ is also real: $\Delta_d= e^{i \theta} (\Delta_1,... , \Delta_5)$ with $\Delta_i$ real, and $\theta$ as an overall phase relative to the phase of the $s$-wave.  The $d$-wave order parameter then becomes finite when its effective quadratic coefficient
\begin{align}
a_d=  a_2 + (q_4  + 2 q_5 \cos 2\theta)\Delta_s ^2  ,
\end{align}
multiplying $\Delta_d^\dagger\Delta_d$ for nonzero $\Delta_s$ in Eq.~(\ref{eq_quartic1}) changes sign. To the leading (second) order in $\lambda$ we therefore find that $a_d$ is lowest for $\theta=0$, when it becomes
\begin{align}
a_d = \frac{8}{5}
\Big [ \zeta \mathcal{N}(\mu,0)  + \frac{2 \kappa a_1}{\zeta} \Big ]
(\frac{\lambda \mu}{T} )^2,
\label{main_a_d}
\end{align}
where $\kappa = 0.0155$ is a constant (see Appendix \ref{append_second_transition}). The quadratic coefficient $a_d$ therefore becomes negative at $T=T_c^d$, with the $d$-wave transition temperature at small parameter $\lambda$ given by (see Appendix \ref{append_second_transition})
\begin{align}
\frac{T_c^d}{T_c^s} = e^{-\frac{\zeta^2}{4\kappa} }
= 0.4807.
\label{TdTc}
\end{align}
At this $d$-wave transition within the $s$-wave state the remaining SO(5) symmetry of the Ginzburg-Landau free energy for $d$-wave components is thus reduced to further to the SO(4).

%================
\section{Summary and Discussion}

To summarize, we considered the Luttinger fermions in the limit of weak spin-orbit, band-inverting, Luttinger parameter, interacting via featureless contact attraction. In this limit the theory becomes almost SU(4) symmetric, and the $s$-wave and the $d$-wave components of the superconducting order parameter together form the six-dimensional irreducible representation. We derived the Ginzburg-Landau free energy both in the limit of vanishing and of weak band-inverting Luttinger parameter, and found that in the latter case the superconducting state at low temperatures has a dominant $i s$ and a subdominant $d$ component, thus breaking the time reversal. In the strict SU(4) limit the superconducting state breaks the symmetry to the SO(5).

Complex order parameter that transforms as an antisymmetric tensor under the group SU(N) has, to the best of our knowledge, so far received only minimal attention in  literature \cite{komarova, wilczek, antonov,Lukas}. One-loop renormalization group study \cite{komarova} finds, for example, only a runaway flow, and suggests that the Ginzburg-Landau free energy in Eq.~(\ref{GL_su4}) leads to the weak first order transition once the order parameter fluctuations are included \cite{herbutbook}. It would be interesting to see if this conclusion would survive higher-order calculations.

We may examine the validity of the criterion for weakness of the Luttinger parameter $\lambda$ in our calculation, namely $\lambda \mu/T<1$, in real semiconductors. Assuming $\lambda\approx 0.1$, and $|\mu| \approx 0.01 eV$ as crude order-of-magnitude estimates, would imply $T_c > 10 K$ for our criterion to be satisfied near the critical temperature, for example. In diamond, for instance, the effective value of $\lambda$ would be somewhat higher \cite{cardona}, $T_c \approx 4 K$ and thus lower \cite{ekimov}, and the value of $|\mu|$ for doped holes less certain \cite{ekimov}, but likely to be also higher. Altogether, in diamond at least it seems likely that assuming $\lambda \mu/T_c >1$ to be more appropriate, and although the Luttinger parameter $\lambda$ itself may be reasonably small, taking into account the energy scales relevant to superconductivity the actual perturbation parameter is not. The coefficients of the Ginzburg-Landau free energy can, on the other hand, in principle be evaluated for all values of $\lambda$, and the interplay of $s$- and $d$-wave order parameters studied that way outside of the perturbative regime considered here. If one assumes the opposite limit $\lambda \mu/T \gg 1$, we expect the $d$-wave critical temperature to likely vanish, and the superconducting state to end up being purely $s$-wave. This is because the $s$-wave, once it develops below its critical temperature, suppresses the $d$-wave components via the quartic term proportional to, presumably positive, coefficient $q_4$ in Eq. (\ref{eq_quartic1}). What is needed to be in the relevant parameter range for the time-reversal broken state we found is therefore a system with higher $T_c$ and low carrier concentration at the same time, which is difficult to achieve \cite{larkin}. We hope the present work will further stimulate the search for superconductivity in lightly doped semiconductors.

\section*{Acknowledgement}
M.Kh. would like to thank S.~Mandal and R.~Boyack for helpful discussions. This work has been supported by the NSERC of Canada.

%===============
\begin{appendix}
\setcounter{figure}{0}
\renewcommand{\thefigure}{C\arabic{figure}}
%==========
\section{SO(6) rotations}
\label{apend_rotation}
We showed that in the SU(4) $\simeq$ SO(6) limit, we can define $\phi$ as an antisymmetric matrix $\phi=-\phi^{\rm T}$, which transform as $\phi \to U\phi U^{\rm T}$ [see Eq.~(\ref{main_phi}) in the main text]. Also,
\begin{align}
U = e^{i \sum_{n=1}^{15} h_n \theta_n},
\end{align}
where $h_n$ represent 15 Hermitian traceless generators of SU(4). They can be chosen to be $\gamma_1$, $\gamma_2$,\dots, $\gamma_5$ and $i\gamma_i \gamma_j$ for $i \neq j$. In this Appendix, we show that the $\gamma_i$ is the generator of the rotations in ``0i" plane while $i\gamma_i \gamma_j$ is a generator of the  rotations in the ``$ij$" plane. Let us show the former explicitly only for $h_1 = \gamma_1$, since the other choices are analogous. In this case, we define
\begin{align}
X &= U \phi U^{\rm T}
\nonumber
\\
&= e^{i \theta_1 \gamma_1} \Big(i\Delta_0 \gamma_{45} + \sum_{a=1}^5 \Delta_a \gamma_a \gamma_{45} \Big) [e^{i \theta_1 \gamma_1}]^{\rm T}.
\end{align}
Note that $\gamma_{1}$, $\gamma_{2}$, and $\gamma_{3}$ are real and Hermitian and hence symmetric while $\gamma_{4}$ and $\gamma_{5}$ are pure imaginary and Hermitian and hence antisymmetric. Therefore,
\begin{align}
X &= e^{i \theta_1 \gamma_1} \Big(i\Delta_0 \gamma_{45} + \sum_{a=1}^5 \Delta_a \gamma_a \gamma_{45} \Big) e^{i \theta_1 \gamma_1}
\nonumber
\\&=
e^{i 2\theta_1 \gamma_1} \Big(i\Delta_0 \gamma_{45} + \Delta_1 \gamma_1 \gamma_{45} \Big) +  \sum_{a=2}^5 \Delta_a \gamma_a \gamma_{45}
\nonumber
\\&=
 \sum_{a=2}^5 \Delta_a \gamma_a \gamma_{45} +
i( \Delta_0 \cos 2\theta_1 + \Delta_1 \sin 2\theta_1 ) \gamma_{45}
\nonumber
\\&\hspace{1cm}
+ (\Delta_1 \cos 2\theta_1  -\Delta_0 \sin 2\theta_1 ) \gamma_{1} \gamma_{45} ,
\end{align}
where we used that $\gamma^2_1 = 1_{4 \times 4}$. This can be viewed as
\begin{equation}
\begin{pmatrix}
\Delta_0 \\ \Delta_1 \\ \Delta_2 \\ \Delta_3 \\ \Delta_4  \\ \Delta_5
\end{pmatrix}
\to
\begin{pmatrix}
\Delta_0 \cos 2\theta_1 + \Delta_1 \sin 2\theta_1 \\
 - \Delta_0\sin 2\theta_1 + \Delta_1  \cos 2 \theta_1
\\ \Delta_2 \\ \Delta_3 \\ \Delta_4  \\ \Delta_5
\end{pmatrix},
\end{equation}
and interpreted as a rotation in the ``01" plane. One can similarly show that all other $\gamma_i$ generate rotations in the ``$0i$" plane.

Let us now take $i\gamma_1 \gamma_2$ as a generator. In this case, we define
\begin{align}
Y &= U \phi U^{\rm T}
\nonumber
\\
&= e^{-\theta \gamma_1 \gamma_2} \Big(i\Delta_0 \gamma_{45} + \sum_{a=1}^5 \Delta_a \gamma_a \gamma_{45} \Big) [e^{-\theta \gamma_1 \gamma_2}]^{\rm T}
\nonumber
\\&=
e^{-\theta \gamma_1 \gamma_2} \Big(i\Delta_0 \gamma_{45} + \sum_{a=1}^5 \Delta_a \gamma_a \gamma_{45} \Big)
e^{\theta \gamma_1 \gamma_2}
\nonumber
\\&=
i\Delta_0 \gamma_{45} + \sum_{a=3}^5 \Delta_a \gamma_a
\gamma_{45}
+
e^{-2\theta \gamma_1 \gamma_2} ( \Delta_1 \gamma_1  + \Delta_2 \gamma_2 ) \gamma_{45}
\nonumber
\\&=
i\Delta_0 \gamma_{45} + \sum_{a=3}^5 \Delta_a \gamma_a \gamma_{45}
+
(\cos 2\theta -\gamma_1 \gamma_2 \sin 2\theta)
\nonumber
\\&\qquad \qquad
\times
( \Delta_1 \gamma_1  + \Delta_2 \gamma_2 ) \gamma_{45} ,
\end{align}
where we used the anticommutation $\gamma_1 \gamma_2 = -\gamma_2 \gamma_1$ in the second line. This can be written as
\begin{equation}
\begin{pmatrix}
\Delta_0 \\ \Delta_1 \\ \Delta_2 \\ \Delta_3 \\ \Delta_4  \\ \Delta_5
\end{pmatrix}
\to
\begin{pmatrix}
\Delta_0 \\
\Delta_1 \cos 2\theta  - \Delta_2 \sin 2\theta  \\
\Delta_1 \sin 2\theta  + \Delta_2 \cos 2\theta \\
\Delta_3 \\ \Delta_4  \\ \Delta_5
\end{pmatrix},
\end{equation}
which clearly is a rotation in the ``12" plane. In exactly the same way one can show that $\gamma_{ij} = i \gamma_i \gamma_j$ generates a rotation in the ``$ij$" plane.

%=========
\section{One-loop integrals and the derivation of the Ginzburg-Landau free energy}
\label{append_loop}
After performing the standard Hubbard-Stratonovich decomposition on the interaction part of the Lagrangian, we get the free energy of the superconducting state per volume as
\begin{align}
f_{\rm sc} = \dfrac{\Delta^{\dagger}  \Delta}{g}
-\frac{1}{\beta} \ln \int \mathcal{D}[\bar{\psi},\psi]
e^{-(S_0 + S_{\rm int}) },
\end{align}
where $\psi_{\bm{p}}(\omega)$ is the Grassmann fermionic field, and $\omega_n = (2n+1)\pi/\beta$ is the fermionic Matsubara frequency for $n\in \mathbb{Z}$. The noninteracting $S_0$ and interacting $S_{\rm int}$ actions are given by
\begin{align}
S_{0} &=\sum_{n} \int \frac{d^3 \bm{p}}{(2\pi)^3}
 \psi^{\dagger}_{\bm{p}}(\omega)
\big[ i\omega_n + \mathcal{H}_0(\bm{p})  \big]
 \psi_{\bm{p}}(\omega) ,
 \nonumber
\\
S_{\rm int} &=\sum_{n,a}\int \frac{d^3 \bm{p}}{(2\pi)^3}
 \big[ \Delta^{\ast}_a~
 \psi_{\bm{p}}^{\rm T}(\omega)
\gamma_{45} \mathcal{A}_a
 \psi_{-\bm{p}}(-\omega) + h.c.  \big],
 \nonumber
\end{align}
where
\begin{equation}
\Delta_a = g \big\langle \psi_{\bm{p}}^{\rm T}(\omega)
\gamma_{45} \mathcal{A}_a
\psi_{-\bm{p}}(-\omega)  \big \rangle,
\label{sc_order}
\end{equation}
is the uniform order parameter, $\gamma_{45} = i \gamma_4 \gamma_5$ is the unitary part of the time-reversal operator, and $\mathcal{A}_a$ are $4 \times 4$ matrices. By integrating out the fermionic field, we get
\begin{align}
f_{\rm sc} &=
 \dfrac{\Delta^{\dagger}  \Delta}{g}
-\frac{1}{\beta} \ln
\Big[
\Big\langle
e^{- S_{\rm int} }
\Big\rangle_{0,{\rm con}}
 \int\mathcal{D}[\bar{\psi},\psi]
e^{-S_0}
\Big]
\nonumber
\\&=
 \dfrac{\Delta^{\dagger}  \Delta}{g}
-\frac{1}{\beta} \ln
(1 + \frac{\langle S^2_{\rm int} \rangle_{0,{\rm con}}}{2!} + \frac{\langle S^4_{\rm int} \rangle_{0,{\rm con}} }{4!} +\dots)
 + f_0,
 \nonumber
\end{align}
where $f_0 = -\frac{1}{\beta} \ln \int \mathcal{D}[\bar{\psi},\psi] e^{-S_0} $ is the normal-state free energy per volume and $\langle \dots \rangle_{0,{\rm con}}$ is the expectation value with respect to $S_0$ over the connected diagrams. Therefore,
\begin{align}
F(\Delta) &= f_{\rm sc}-f_0 \approx
\dfrac{\Delta^{\dagger}  \Delta}{g}
-\frac{1}{2!\beta}  \sum_{n}\int \frac{d^3 \bm{p}}{(2\pi)^3}
 \Big\langle  S^2_{\rm int} \Big\rangle_{0,{\rm con}}
 \nonumber
\\&
\hspace{1.8cm}
-\frac{1}{4!\beta}
\sum_{n}\int \frac{d^3 \bm{p}}{(2\pi)^3}
\Big\langle  S^4_{\rm int} \Big\rangle_{0,{\rm con}}  + \dots,
\nonumber
\end{align}
where we used $\log(1+x) \approx x$ since the superconducting order parameter $\Delta_a$ is small close to the critical temperature, so $S_{\rm int}$ can be treated as a perturbation. After performing the one-loop integral and keeping the fourth-order terms in $\Delta$, we obtain
\begin{align}
F(\Delta)&= \dfrac{\Delta^{\dagger}  \Delta}{g}
-\frac{2}{ \beta}
\sum_n \int {\rm tr}\Big[
G_0(-Q)
\mathcal{A}^{\dagger}_a
G_0(Q)
\mathcal{A}_b
 \Big]
\Delta_a \Delta^{\ast}_b
\nonumber
\\&\quad
+ \frac{4}{\beta}
\sum_n \int
{\rm tr}\Big[
G_0(-Q)
\mathcal{A}^{\dagger}_a
G_0(Q)
\mathcal{A}_b
G_0(-Q)
\mathcal{A}^{\dagger}_c
\nonumber
\\&\hspace{1.5cm}
\times
G_0(Q)
\mathcal{A}_d
\Big]
\Delta_a \Delta^{\ast}_b \Delta_c \Delta^{\ast}_d,
\end{align}
where we introduced $Q = (\bm{p},\omega_n)$ for brevity and employed
$\gamma_{45} G^{\rm T}(Q)\gamma_{45} = G(Q)$. Therefore, we get
$F(\Delta) = F_2(\Delta)  + F_4(\Delta) $ where
\begin{align}
F_2(\Delta) &= \frac{\Delta^{\dagger}  \Delta}{g} + \sum_{a,b} K_{ab} \Delta_a \Delta^{\ast}_b,
\label{quad}
\\
F_4(\Delta) &=
\sum_{a,b,c,d}
 K_{abcd}
\Delta_a \Delta^{\ast}_b
\Delta_c \Delta^{\ast}_d,
\label{quar}
\end{align}
and
\begin{widetext}
\begin{align}
K_{ab} &= -\frac{2}{\beta}  \sum_{n=-\infty}^{\infty}
\int
\hspace{-1mm}
\frac{d^3 \bm{p}}{(2\pi)^3}
{\rm tr}
\big[
G_0(-Q) \mathcal{A}^{\dagger}_a G_0(Q) \mathcal{A}_b \big],
\\
K_{abcd} &=\frac{4}{\beta}  \sum_{n=-\infty}^{\infty}
\int
\hspace{-1mm}
\frac{d^3 \bm{p}}{(2\pi)^3}
{\rm tr} \big[
G_0(-Q) \mathcal{A}^{\dagger}_a
G_0(Q) \mathcal{A}_b
G_0(-Q) \mathcal{A}^{\dagger}_c
G_0(Q) \mathcal{A}_d
\big].
\label{eq_kabcd}
\end{align}
\end{widetext}
The free propagator is defined as
\begin{align}
G_0(Q) &=
\big[ i \omega_n 1_{4 \times 4}  - \mathcal{H}_0(\bm{p}) \big]^{-1},
\end{align}
where $1_{4 \times 4} $ is $4 \times 4$ unit matrix. It therefore reads
\begin{align}
G_0(Q) =
\frac{(i \omega_n  - p^2 + \mu) 1_{4 \times 4} + \lambda d_a(\bm{p}) \gamma_a}{(i \omega_n  - p^2 + \mu)^2 - \lambda^2 p^4}.
\end{align}

%=========
\section{Calculation of the Ginzburg-Landau coefficients in the SU(4) limit when $\lambda = 0$}
\label{append_lambda_zero}

In the SU(4) limit, the mean-field Ginzburg-Landau free energy given by Eq.~(\ref{GL_su4}) in the main text can be rewritten as $F(\Delta) = F_2(\Delta) + F_4(\Delta)$ where
\begin{align}
F_2(\Delta) &= 4a \Delta ^\dagger \Delta,
\label{f2_su4}
\\
F_4(\Delta) &=(16b_1 + 8b_2) (\Delta ^\dagger \Delta )^2
-4 b_2 |\Delta ^{\rm T} \Delta|^2.
\label{f4_su4}
\end{align}
To determine the three unknown coefficients $a$, $b_1$, and $b_2$, we use two distinct normalized configurations $\Delta^{s,d}_1 = (1,0,0,0,0,0)$ and $\Delta^{s,d}_2 = \frac{1}{\sqrt{2}}(1,i,0,0,0,0)$ and insert them in Eqs.~(\ref{f2_su4}) and (\ref{f4_su4}) as well as Eqs.~(\ref{quad}) and (\ref{quar}). This results in three linearly independent equations
\begin{align}
\frac{4}{g} -2T\sum_n \hspace{-1mm}
 \int
 \hspace{-2mm}  \frac{d^3 \bm{p}}{(2\pi)^3}
\frac{4}{(p^2 - \mu)^2 + \omega_n^2} =
F_2(\Delta^{s,d}_1) =  4 a,
\nonumber
\\
4T \sum_n \hspace{-1mm}
\int
 \hspace{-2mm}  \frac{d^3 \bm{p}}{(2\pi)^3}
\frac{4}{\big[(p^2 - \mu)^2 + \omega_n^2 \big]^2} =
F_4(\Delta^{s,d}_1) = 16b_1 + 4b_2,
\nonumber
\\
4T \sum_n \hspace{-1mm}
\int
\hspace{-2mm} \frac{d^3 \bm{p}}{(2\pi)^3}
\frac{8 }{\big[(p^2 - \mu)^2 + \omega_n^2 \big]^2} =
F_4(\Delta^{s,d}_2) = 16b_1 + 8b_2.
\nonumber
\end{align}
Solving these equations then yields
\begin{align}
a &=
\frac{1}{g} - 2T \hspace{-1.5mm} \sum_{n=-\infty}^{\infty}
\int
\frac{d^3 \bm{p}}{(2\pi)^3}
\frac{1}{(p^2 - \mu)^2 + \omega_n^2},
\label{appA_a}
\\
b_1 &= 0,
\\
b_2 &= 4T \hspace{-1.5mm}  \sum_{n=-\infty}^{\infty}
\int
\frac{d^3 \bm{p}}{(2\pi)^3}
\frac{1}{\big[(p^2 - \mu)^2 + \omega_n^2 \big]^2}.
\label{appA_b}
\end{align}
To simplify Eqs.~(\ref{appA_a}) and (\ref{appA_b}) further, we first introduce $\xi = p^2 - \mu$ and then use
\begin{align}
 \int
\frac{d^3 \bm{p}}{(2\pi)^3}
=
\frac{1}{4}
\int_{-\Omega}^{\Omega}
\mathcal{N}(\mu, 0)
 d\xi.
\end{align}
Performing the finite temperature Matsubara summation and then introducing $x =\xi/T$ gives
\begin{align}
a &=
\frac{1}{g} -
\frac{\mathcal{N}(\mu,0)}{2}
\int_{0}^{\Omega/T}
\frac{dx}{x} \tanh \frac{x}{2},
\\
b_2 &=
\frac{\mathcal{N}(\mu,0)}{2 T^2}
\int_{0}^{\Omega/T}
\frac{\sinh x -x}{x^3 (1+\cosh x )} dx.
\end{align}
At low temperatures, the upper limit of both integrals is large so we  approximate it as $\Omega/T \to \infty$. The first integral diverges logarithmically at low temperatures and after integration by parts one finds
\begin{align}
a &= \frac{1}{g} -
\frac{\mathcal{N}(\mu,0)}{2}
(\log \frac{\Omega}{T} + C),
\label{appA_log}
\\
b_2 &=
\frac{\zeta \mathcal{N}(\mu,0)}{2 T_c ^2},
\label{eq_b2}
\end{align}
where
\begin{align}
\zeta = \int_{0}^{\infty}
\frac{\sinh x -x}{x^3 (1+\cosh x )} dx = 0.2131,
\label{eq_zeta}
\\
C = -\frac{1}{2}\int_0^{\infty}  \log(x) ~{\rm sech}^2(\frac{x}{2}) ~dx
= \log \frac{2e^{\gamma}}{\pi} ,
\end{align}
and $\gamma = 0.5772$ is known as Euler's constant. Equations~(\ref{appA_log}) and (\ref{eq_b2}) correspond to Eqs.~(\ref{eq_log}) and (\ref{eq_b2_main}) in the main text, respectively.

%=========
\section{Calculation of the Ginzburg-Landau coefficients when $\lambda \neq 0$}
\label{append_lambda_finite}

In this case, we have to treat $s$- and $d$-wave order parameters as two distinct objects. The Ginzburg-Landau free energy up to quartic order given by Eq.~(\ref{eq_quartic1}) in the main text can be rewritten as
\begin{align}
F(\Delta_s,\Delta_d) &= F_2(\Delta_s,\Delta_d) + F_4(\Delta_s,\Delta_d),
\end{align}
where
\begin{align}
F_2(\Delta_s,\Delta_d) &= a_1 |\Delta_s|^2 + a_2 \Delta_d ^\dagger \Delta_d,
\label{appC_f2}
\\
F_4(\Delta_s,\Delta_d) &=
q_1 |\Delta_s |^4
+ q_2 (\Delta_d ^\dagger \Delta_d) ^2
\nonumber
\\&\quad
+ q_3 |\Delta_d ^{\rm T} \Delta_d|^2
+ q_4 |\Delta_s|^2  \Delta_d ^\dagger \Delta_d
\nonumber
\\&\quad
+ q_5 (\Delta^{\ast 2}_s \Delta_d ^{\rm T} \Delta_d  +  \Delta^{2}_s \Delta_d ^\dagger \Delta_d ^* )
\nonumber
\\&\quad
+ q_6 \Delta^{\ast}_s {\rm tr} (M^2 M^{\ast} ) +  q^{\ast}_6 \Delta _s {\rm tr} (M^{\ast 2} M )
\nonumber
\\&\quad
+ q_7 {\rm tr}(M M^{\ast} M M^{\ast}).
\label{appC_f4}
\end{align}
Here,
\begin{equation}
M = \sum_{a=1}^5 \Delta_a M_a,
\end{equation}
where the five Gell-Mann matrices
\begin{align}
M_1 &=
\begin{pmatrix}
1 & 0 & 0 \\
0 & -1 & 0 \\
0 & 0 &0
\end{pmatrix},
\quad
M_2 = \frac{1}{\sqrt{3}}
\begin{pmatrix}
-1 & 0 & 0 \\
0 & -1 & 0 \\
0 & 0 & 2
\end{pmatrix} ,
\nonumber
\\
M_3 &=
\begin{pmatrix}
0 & 0 & 1 \\
0 & 0 & 0 \\
1 & 0 &0
\end{pmatrix},
\quad
M_4 =
\begin{pmatrix}
0 & 0 & 0 \\
0 & 0 & 1 \\
0 & 1 &0
\end{pmatrix} ,
\nonumber
\\
M_5 &=
\begin{pmatrix}
0 & 1 & 0 \\
1 & 0 & 0 \\
0 & 0 & 0
\end{pmatrix} ,
\label{Gmann}
\end{align}
provide a basis of three-dimensional real traceless symmetric matrices.

In order to calculate $a_1$ and $a_2$, we define $\Delta_1 = (1,0,0,0,0)$ and employ Eqs.~(\ref{quad}) and (\ref{appC_f2}). This yields two linearly independent equations $F_2(1,0) = a_1$ and $F_2(0,\Delta_1) = a_2$. Solving these equations gives
\begin{widetext}
\begin{align}
a_1 &= \frac{4}{g}-8T
\sum_{n=-\infty}^{\infty}
 \int
\frac{d^3 \bm{p}}{(2\pi)^3}
\frac{\lambda^2 p^4 +\left(p^2-\mu \right)^2 + \omega_n^2}{
\big[\big(p^2(1+\lambda) - \mu \big)^2 + \omega_n^2
\big]
\big[ \big(p^2(1-\lambda) - \mu \big)^2 + \omega_n^2
\big]},
\label{appC_a1}
\\
a_2 &= \frac{4}{g}-8T
 \sum_{n=-\infty}^{\infty}
 \int
\frac{d^3 \bm{p}}{(2\pi)^3}
\frac{ -\dfrac{3}{5} \lambda^2 p^4 +\left(p^2-\mu \right)^2 + \omega_n^2}{
\big[\big(p^2(1+\lambda) - \mu \big)^2 + \omega_n^2
\big]
\big[\big(p^2(1-\lambda) - \mu \big)^2 + \omega_n^2
\big]}.
\label{appC_a2}
\end{align}
\end{widetext}
If we define $\xi_{\pm} = p^2(1 \pm \lambda) - \mu$, the integrand of Eq.~(\ref{appC_a1}) can be written as

\begin{align}
\hspace{-2mm}
\frac{\lambda^2 p^4 +\left(p^2-\mu \right)^2 + \omega_n^2}{
(\xi_{+}^2 + \omega_n^2  )
(\xi_{-}^2 + \omega_n^2)}
&=
\frac{1}{2}
\frac{\xi^2_{+} + \xi^2_{-} + 2\omega_n^2}{
(\xi_{+}^2 + \omega_n^2  )
(\xi_{-}^2 + \omega_n^2)}
\nonumber
\\&=
\frac{1}{2} \Big[
 \frac{1}{\xi^2_{+} + \omega_n^2 }
 +\frac{1}{\xi^2_{-} + \omega_n^2}
 \Big].
\end{align}
By using
\begin{align}
 \int
\frac{d^3 \bm{p}}{(2\pi)^3}
=
\frac{1}{2}
\int_{-\Omega}^{\Omega}
\mathcal{N}_{\eta}(\mu, \lambda)
 d\xi_{\eta},
\end{align}
where $\mathcal{N}_{\eta}(\mu, \lambda)$ is defined by Eq.~(\ref{DOS}) in the main text, we get
\begin{align}
 a_1 = \frac{4}{g} -2T
\mathcal{N}(\mu, \lambda)
\sum_n
\int_{-\Omega}^{\Omega}
\frac{d\xi}{\xi^2 + \omega^2_n }.
\nonumber
\end{align}
After performing the Matsubara summation, we get
\begin{align}
a_1 = \frac{4}{g} - 2 \mathcal{N}(\mu,\lambda)
\int_{0}^{\Omega/T}
\frac{dx}{x}
\tanh\frac{x}{2},
\end{align}
where we introduced $x= \xi/T$ after performing the Matsubara summation. Calculating the integral by parts yields
\begin{align}
a_1 =  \frac{4}{g}
- 2 \mathcal{N}(\mu,\lambda)
(\log \frac{\Omega}{T} + C).
\label{appC_loga1}
\end{align}
Furthermore, to simplify Eq.~(\ref{appC_a2}), we write it as
\begin{align}
\hspace{-3mm}
a_2 &= a_1 +\frac{64T}{5}
\sum_{n} \int
\frac{d^3 \bm{p}}{(2\pi)^3}
\frac{\lambda^2 p^4 }{
(\xi_{+}^2 + \omega_n^2  )
(\xi_{-}^2 + \omega_n^2)},
\end{align}
so $a_2>a_1$ at finite $\lambda$ for all temperatures. For small $\lambda$, the integrand can be expanded around $\lambda = 0$ as

\begin{align}
a_2 &= a_1 +\frac{64T}{5}
\sum_{n} \int
\frac{d^3 \bm{p}}{(2\pi)^3}
\frac{(\lambda \mu)^2 }{
(\xi^2 + \omega_n^2 )^2} + \mathcal{O}((\lambda \mu)^4),
\nonumber
\end{align}
where we approximate $p^2 \approx \mu $. After performing the Matsubara summation we get
\begin{align}
a_2 = a_1 + \frac{8 \zeta}{5} \mathcal{N}(\mu,0) (\frac{\lambda \mu}{T})^2
+ \mathcal{O}((\lambda \mu)^4),
\label{diff_a_2_1}
\end{align}
which is Eq.~(\ref{maintext_diff_a_2_1}) in the main text.

\begin{table}[!t]
\centering
\includegraphics[scale= 0.39]{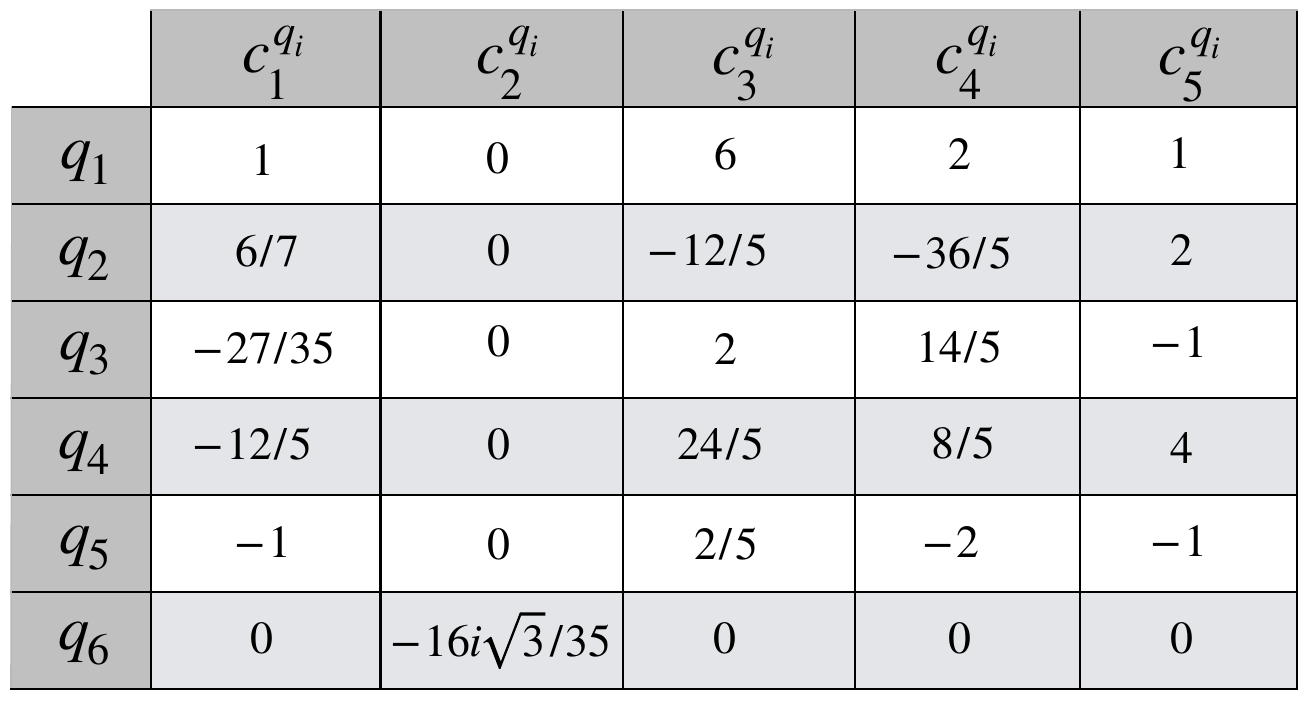}
\caption{The values of $c^{q_i}_1,c^{q_i}_2,\dots,c^{q_i}_5$
corresponding to each $q_i$ in Eq.~(\ref{eq_q_i})
for $i=1,2,\dots,6$ at $\lambda \neq 0$.}
\label{table1}
\end{table}

In order to calculate the seven quartic coefficients $q_i$, we introduce three more distinct normalized $d$-wave configurations
$\Delta_2 = \frac{1}{\sqrt{2}}(1,i,0,0,0)$,
$\Delta_3 = \frac{1}{\sqrt{2}}(1,0,i,0,0)$, and
$\Delta_4 = \frac{i}{\sqrt{2}}(1,0,1,0,0)$. Substitute them in Eqs.~(\ref{quar}) and (\ref{appC_f4}) results in
\begin{align}
F_4(1,0) &= q_1,
\nonumber
\\
F_4(0,\Delta_1) &= q_2 + q_3 + 2q_7,
\nonumber
\\
F_4(0, \Delta_2) &= q_2 + \frac{4}{3} q_7,
\nonumber
\\
F_4(0,\Delta_3) &= q_2 + 2 q_7,
\nonumber
\\
F_4(1,\Delta_1) &= q_1 + q_2 + q_3 + q_4 + 2 q_5 + 2 q_7,
\nonumber
\\
F_4(1,\Delta_2) &= q_1 + q_2 + q_4 + \frac{4}{3} q_7,
\nonumber
\\
F_4(1,\Delta_4) &=
q_1 + q_2 + q_3 + q_4 - 2 q_5 + \frac{3i}{2\sqrt{2}} (q_6 - q^{\ast}_6) + 2 q_7.
\nonumber
\end{align}
After solving these equations for the seven unknown $q_i$, one finds
\begin{widetext}
\begin{align}
q_i = 16 T \sum_{n=-\infty}^{\infty}
\int
\frac{d^3 \bm{p}}{(2\pi)^3}
\frac{
c^{q_i}_1 \lambda^4 p^8 + c^{q_i}_2 \lambda^3 p^6 (p^2 - \mu)
+ \lambda ^2 p^4 \Big[ c^{q_i}_3 \left(p^2-\mu\right)^2 + c^{q_i}_4 \omega_n^2\Big] + c^{q_i}_5 \Big[\left(p^2-\mu\right)^2 + \omega_n^2\Big]^2
}{
\Big[\big(p^2(1+\lambda) - \mu \big)^2+\omega_n^2
\Big]^2\Big[\big(p^2(1-\lambda) - \mu \big)^2 + \omega_n^2 \Big]^2},
\label{eq_q_i}
\end{align}
\end{widetext}
where the values of $c^{q_i}_i$ are given in Table \ref{table1} for $i = 1,2,\dots,5$ . Note that the coefficient $q_6\sim \lambda^3$, from the last row in the table.

%=========
\section{Second phase transition from $s$-wave to time-reversal breaking $is+d$-wave superconducting phase}
\label{append_second_transition}

For $T< T_c^s$, the $d$-wave order parameter becomes finite when its effective quadratic coefficient
\begin{align}
a_d &=  a_2 + (q_4  + 2 q_5 \cos 2\theta)\Delta_s ^2
\nonumber
\\&
=a_2 - \frac{a_1}{2q_1}(q_4 + 2q_5 \cos 2\theta),
\label{a_dd}
\end{align}
changes sign. In order to simplify $a_d$, we find that to the leading order in $\lambda$ the denominator of Eq.~(\ref{eq_q_i}) can be expanded as
\begin{align}
\frac{1}{(\xi^2_{+} + \omega^2)^2(\xi^2_{-} + \omega^2)^2}
&\approx
\frac{1}{(\xi^2 + \omega^2)^4}
\Big[1 + \frac{4(\xi^2 - \omega^2)}{(\xi^2 + \omega^2)^2}
(\lambda p^2)^2
\nonumber
\\&\qquad
+ \mathcal{O}((\lambda p^2)^4) \Big],
\end{align}
where $\xi_{\pm} = p^2(1 \pm \lambda) - \mu$ and $\xi = p^2 - \mu$. Therefore, we can approximate $q_1$, $q_4$, and $q_5$ as
\begin{align}
q_i &\approx
c_5^{q_i}  q_{10} +
16T \sum_n \int \frac{d^3 \bm{p}}{(2\pi)^3}
\Big[
\frac{c_3^{q_i} (p^2 - \mu)^2 + c_4^{q_i} \omega_n^2}{(\xi^2 + \omega_n^2)^4}
\nonumber
\\&\qquad
+ \frac{4 c_5^{q_i} (\xi^2 - \omega^2)}{(\xi^2 + \omega^2)^4}
\Big]
(\lambda \mu)^2,
\label{appE_qi}
\end{align}
where we used $p^2 \approx \mu$, and define $q_{10} = q_1(\lambda = 0)$. Note that performing the integral close to the Fermi surfaces for the term whose coefficient is $c_2^{q_i}$ inside the integrand of Eq.~(\ref{eq_q_i}) gives zero as the corresponding integrand is odd with respect to $\xi$. Also, the integral over the second line of Eq.~(\ref{appE_qi}) becomes zero at $T=0$, so it is parametrically smaller at small temperatures, so we neglect it. Therefore, we obtain
\begin{align}
\frac{q_4 + 2q_5}{2q_1} \approx
1- \frac{16r}{5 } (\lambda \mu)^2,
\end{align}
where we set $\theta = 0$ to make $a_d$ minimal, and defined the ratio
\begin{align}
r &= \Big( T\sum_n  \hspace{-1mm}
\int_{-\Omega}^{\Omega}
 \frac{d \xi}{(\xi^2 +\omega_n^2)^3} \Big)
 \Big(T\sum_n \hspace{-1mm}
\int_{-\Omega}^{\Omega}
 \frac{d\xi}{(\xi^2 +\omega_n^2)^2} \Big)^{-1}.
\nonumber
\end{align}
Simplifying this further gives
\begin{align}
\frac{q_4 + 2q_5}{2q_1} \approx
1- \frac{16}{5 } (\frac{\lambda \mu}{T})^2
(\frac{\kappa}{\zeta}),
\label{eq_ratio}
\end{align}
at low temperatures where
\begin{align}
\kappa &= 2T^5
\sum_n \int_{-\Omega}^{\Omega}
 \frac{d\xi}{(\xi^2 +\omega_n^2)^3}
\nonumber
\\&
=
\int_0^{\infty}
\frac{3 (\sinh x - x) - x^2 \tanh \frac{x}{2} }{4 x^5 (1+\cosh x)} ~dx
= 0.0155,
\end{align}
and
\begin{align}
\zeta&= 2T^3
\sum_n  \int_{-\Omega}^{\Omega}
\frac{d\xi}{(\xi^2 +\omega_n^2)^2} = 0.2131,
\end{align}
as already defined by Eq.~{(\ref{eq_zeta})}. Substituting Eq.~(\ref{eq_ratio}) in the second line of Eq.~(\ref{a_dd}) at $\theta = 0$ yields
\begin{align}
a_d &=
a_2 - a_1 + \frac{16 a_1}{5}
(\frac{\lambda \mu}{T})^2
(\frac{\kappa}{\zeta}).
\label{eq_a_d}
\end{align}
By employing Eq.~(\ref{diff_a_2_1}), we finally get
\begin{align}
a_d = \frac{8}{5}
\Big [ \zeta \mathcal{N}(\mu,0)  + \frac{2 \kappa a_1}{\zeta} \Big ]
(\frac{\lambda \mu}{T} )^2,
\end{align}
which is Eq.~(\ref{main_a_d}) in the main text. Substituting
Eq.~(\ref{appC_loga1}) for $a_1$ and then using the fact that below $T^{s}_c$
\begin{align}
\frac{4}{g} =
2 \mathcal{N}(\mu,\lambda)
(\log \frac{\Omega}{T^s_c} + C),
\end{align}
we get
\begin{align}\
a_d = \frac{8}{5}
\mathcal{N}(\mu,0)
\Big [\zeta + \frac{4\kappa}{\zeta}  \log (\frac{T}{T_c^s}) \Big ]
(\frac{\lambda \mu}{T} )^2,
\end{align}
to the leading order in $\lambda$. After even further simplification, we find that $a_d$ changes sign at $T_c^d$ where
\begin{align}
 \frac{T_c^d}{T_c^s} = e^{-\frac{\zeta^2}{4\kappa} }
= 0.4807,
\end{align}
which is Eq.~(\ref{TdTc}) in the main text.

\end{appendix}

%===============


\begin{thebibliography}{99}


\bibitem{luttinger} J. M. Luttinger, Quantum theory of cyclotron resonance in semiconductors: general theory, Phys. Rev. {\bf 102}, 1030 (1956).


\bibitem{butch} N. P. Butch, P. Syers, K. Kirshenbaum, A. P. Hope, and J. Paglione, Superconductivity in the topological semimetal YPtBi, Phys. Rev. B {\bf 84}, 220504(R) (2011).

\bibitem{bay} T. V. Bay,  M. Jackson, C. Paulsen, C. Baines, A. Amato, T. Orvis, M.C. Aronson, Y.K. Huang, and A. de Visser, Low field magnetic response of the non-centrosymmetric superconductor YPtBi, Sol. St. Comm. {\bf 183}, 13 (2014).

\bibitem{kim} H. Kim, K. Wang, Y. Nakajima, R. Hu, S. Ziemak, P. Syers, L. Wang L, H. Hodovanets, J. D. Denlinger, P. M. R. Brydon, D. F. Agterberg, M. A. Tanatar, R. Prozorov, and J. Paglione, Beyond triplet: unconventional superconductivity in a spin-3/2 topological semimetal, Sc. Adv. {\bf 4}, eaao4513 (2018).

\bibitem{boettcher1}I. Boettcher and I. F. Herbut, Superconducting quantum criticality in three-dimensional Luttinger semimetals, Phys. Rev. B {\bf 93}, 205138 (2016).

\bibitem{meinert} M. Meinert, Unconventional superconductivity in YPtBi and related topological semimetals, Phys. Rev. Lett. {\bf 116}, 137001 (2016).

\bibitem{brydon} P. M. R. Brydon, L. Wang, M. Weinert, and D. F. Agterberg, Pairing of
j=3/2 fermions in Half-heusler superconductors, Phys. Rev. Lett. {\bf 116}, 177001 (2016).


\bibitem{savary} L. Savary, J. Ruhman, J. W. F. Venderbos, Superconductivity in three-dimensional spin-orbit coupled semimetals,  L. Fu, and P. A. Lee, Phys. Rev. B {\bf 96}, 214514 (2017).

\bibitem{boettcher2} I. Boettcher and I. F. Herbut, Unconventional superconductivity in Luttinger semimetals: theory of complex tensor order and the emergence of the uniaxial nematic state, Phys. Rev. Lett. {\bf 120}, 057002 (2018).

\bibitem{venderbos} J. W. F. Venderbos, L. Savary, J. Ruhman, P. A. Lee, and L. Fu, Pairing states of spin-3/2 fermions: symmetry-enforced topological gap functions, Phys. Rev. X {\bf 8}, 011029 (2018).

\bibitem{roy} B. Roy, S. A. A. Ghorashi, M. S. Foster, and A. H. Nevidomskyy, Topological superconductivity of spin-3/2 carriers in a three-dimensional doped Luttinger semimetal, Phys. Rev. B {\bf 99}, 054505 (2019).

\bibitem{mandal} I. F. Herbut, I. Boettcher, and S. Mandal, Ground state of the three-dimensional BCS $d$-wave superconductor, Phys. Rev. B {\bf 100}, 104503 (2019).

\bibitem{link} J. Link and I. F. Herbut, $p$-wave superconductivity in Luttinger semimetals, Phys. Rev. B {\bf 105}, 134522, (2022).

\bibitem{Ipsita} I. Mandal, Fate of superconductivity in three-dimensional disordered Luttinger semimetals, Ann. of Phys. 392, 179 (2018).

\bibitem{ghorashi} S. A. A. Ghorashi, P. Hosur, and C.-S. Ting, Irradiated three-dimensional Luttinger semimetal: a factory for engineering Weyl semimetals, Phys. Rev B {\bf 97}, 205402 (2018).

\bibitem{cardona} P. Y. Yu and M. Cardona, {\sl Fundamentals of semiconductors}, (Springer, Heidelberg, 1996).

\bibitem{ekimov} E. A. Ekimov, V. A. Sidorov, E. D. Bauer, N. N. Mel'nik, N. J. Curro, J. D. Thompson, and S. M. Stishov, Superconductivity in diamond, Nature {\bf 428}, 542 (2004).


\bibitem{herbutfisherbook} I. F. Herbut, preprint arXiv:2304.07654, to appear in {\sl 50 years of renormalization group/Dedicated to the memory of Michael E. Fisher}, ed. by A. Aharony, O. Entin-Wohlman, D. Huse, and L. Radzihovsky, (World Scientific, Singapore, 2024).


\bibitem{herbutmandal} I. F. Herbut and S. Mandal, SO(8) unification and the large-$N$ theory of superconductor-insulator transition of two-dimensional Dirac fermions, Phys. Rev. B {\bf 108}, L161108 (2023).

\bibitem{zee} A. Zee, {\sl Group Theory in Nutshell for Physicists}, (Princeton University Press, Princeton, 2016).

\bibitem{abrikosov} A. A. Abrikosov, Calculation of critical indices for zero-gap semiconductors, Sov. Phys. JETP {\bf 66}, 1443 (1974).


\bibitem{moon} E.-G. Moon, C. Xu, Y. B. Kim, and L. Balents, Non-Fermi-liquid and topological states with strong spin-orbit coupling, Phys. Rev. Lett. {\bf 111}, 206401 (2013).

\bibitem{herbut1} I. F. Herbut and L. Janssen, Topological Mott insulator in three-dimensional systems with quadratic band touching, Phys. Rev. Lett. {\bf 113}, 106401 (2014).
    
\bibitem{janssen1} L. Janssen and I. F. Herbut, Nematic quantum criticality in three-dimensional Fermi system with quadratic band touching, Phys. Rev. B {\bf 92}, 045117 (2015); Excitonic instability of three-dimensional gapless semiconductors: Large-$N$ theory, Phys. Rev. B {\bf 93}, 165109 (2016).  


\bibitem{boettcher3} I. Boettcher and I. F. Herbut,  Anisotropy induces non-Fermi-liquid behavior and nematic magnetic order in three-dimensional Luttinger semimetals, Phys. Rev. B {\bf 95}, 075149 (2017).

\bibitem{herbut2} I. F. Herbut, Hidden role of antiunitary operators in Fierz transformation, Phys Rev. D {\bf 100}, 116015 (2019).

\bibitem{sim} G.-B. Sim, A. Mishra, M. J. Park, Y. B. Kim, G. Y. Cho, S.-B. Lee, Topological
$d+s$ wave superconductors in a multiorbital quadratic band touching system, Phys. Rev. B {\bf 100}, 064509 (2019).

\bibitem{wilczek} See, R. B. Pisarski and F. Wilczek, Remarks on the chiral phase transition in chromodynamics, Phys. Rev. D {\bf 29}, 338(R) (1984), for complex matrix order parameter that transform under SU(N) $\times$ SU(M) relevant to QCD.


\bibitem{antonov} See, N. V. Antonov, M. V. Kompaniets, N. M. Lebedev, Critical behaviour of the $O(n)-\phi^4$ model with an antisymmetric tensor order parameter, J. Phys. A: Math. Theor. {\bf 46}, 405002 21013), for the Ginzburg-Landau free energy of a real anisymmetric tensor order parameter.

\bibitem{Lukas} L. Janssen and U. F. P. Seifert, Phase diagrams of SO(N) Majorana-Hubbard models: dimerization, internal symmetry breaking, and fluctuation-induced first-order transitions, Phys. Rev. B {\bf 105}, 045120 (2022).

\bibitem{komarova} M. V. Komarova, M. Yu. Nalimov, and J. Honkonen, Temperature Green's functions in Fermi systems: the superconducting phase transition, Th. and Math. Phys., {\bf 176}(1), 906 (2013).

\bibitem{herbutbook} I. Herbut, {\sl A Modern Approach to Critical Phenomena}, (Cambridge University Press, Cambridge, England, 2007), Ch. 4.

\bibitem{larkin} V. L.  Gurevich, A. I. Larkin, Yu. A. Firsov, Possibility of superconductivity in semiconductors, Sov. Phys. Solid State {\bf 4}, 131 (1962).




\end{thebibliography}
\end{document}